\documentclass{article}
\usepackage[utf8]{inputenc}
\usepackage{multicol} 
\usepackage{graphicx}
\usepackage{indentfirst} 
\usepackage{booktabs} 
\usepackage{amsmath}
\usepackage{float} 
\usepackage{rotating} 
\usepackage{caption} 
\usepackage{subcaption} 
\usepackage{fullpage} 
\usepackage[english]{babel} 
\usepackage{csquotes}
\usepackage{listings} 
\usepackage[dvipsnames]{xcolor} 
\usepackage{verbatim} 
\usepackage{lipsum} 
\usepackage{upgreek}  
\usepackage{tikz} 
\usepackage{xfrac} 
\usepackage{multirow} 
\usepackage{titlesec} 
\usepackage[sorting=none,style=nature,citestyle=numeric-comp]{biblatex}
\bibliography{references.bib}
\usepackage{url}
\usepackage{subfiles}
\usepackage[toc,page]{appendix}
\usepackage{tabularx}
\usepackage{amsmath,amssymb,lmodern}
\usepackage{mathrsfs}
\usepackage{epsfig}
\usepackage{enumitem} 
\usepackage{outlines} 

\numberwithin{equation}{section}
\numberwithin{figure}{section}

\titlespacing\section{0pt}{12pt plus 4pt minus 2pt}{0pt plus 2pt minus 2pt}
\titlespacing\subsection{0pt}{12pt plus 4pt minus 2pt}{0pt plus 2pt minus 2pt}
\titlespacing\subsubsection{0pt}{12pt plus 4pt minus 2pt}{0pt plus 2pt minus 2pt}

\title{
\begin{center}
\large{\textbf{Zero-field deterministic all-optical writing and annihilation of nanometer-scale skyrmion bubbles}}\\
\fontsize{5mm}{7mm}\selectfont 
\end{center}
}

\author{M.G. van der Schans, W.P.M. de Kleijne, M.A. Brozius, B. Koopmans}

\begin{document}
\maketitle

Skyrmions are highly stable chiral magnetic spin textures with non-trivial topology. They can act as quasi-particles that can be generated, manipulated and annihilated, and hold promise for future memory and logic devices. As of now, all-optical stochastic nucleation of skyrmion ensembles, mostly in small applied magnetic fields, has been shown. However, to research their true potential, the ability to selectively toggle switch individual single skyrmions would be highly beneficial. In this paper, we demonstrate the field-free optical control of single stable skyrmions via single femtosecond laser pulses with diameters down to 175 nm, containing a fixed chirality. By engineering ferrimagnetic Co/Gd-based multilayers, we resolve the competition between deterministic and stochastic processes, and thereby overcome the challenge of optically writing and annihilating sub-micron skyrmions on demand. Our work is envisioned to fuel applications of skyrmion-based applications and opens up further endeavors in research related to the behaviour of more complicated skyrmion-based textures.

\newpage
\begin{multicols}{2}
\section*{Introduction}
    \label{Sections:Introduction}
    Skyrmions are chiral spin textures which, due to their non-trivial topology, offer great potential for applications in numerous computational fields \cite{li2021magnetic,khodzhaev2025voltage,psaroudaki2023skyrmion,beneke2024gesture,PhysRevB.109.174420,tomasello2014strategy}. Specifically within magnetic thin film research, the non-trivial topology is typically induced via the interfacial Dzyaloshinskii-Moriya Interaction (DMI) \cite{PhysRevB.94.064413,moreau2016additive,soumyanarayanan2017tunable}. For larger diameters, skyrmions are typically further stabilized via dipolar interactions. A commonly proposed application for skyrmions is the skyrmion racetrack \cite{tomasello2014strategy,zelent2023stabilization,belrhazi2022nucleation,he2023all}, where skyrmions are driven by a current that exerts a spin-transfer or spin-orbit torque. Here, the existence --- or absence --- of a skyrmion is defined as a 1 or 0-bit respectively, thus allowing for classic logic operations. There are, however, other  endeavors beyond the skyrmion-racetrack. Specifically, skyrmion-skyrmion interactions in skyrmion-based lattices or textures can potentially give rise to interesting exotic behaviour \cite{PhysRevLett.119.077204,gruber2025real,kato2023topological}. Progress in this exciting research field would highly profit from novel tools to deterministically create and annihilate single, fully mobile, skyrmions on demand, and preferably  without the need of an externally applied stabilizing magnetic field.

Numerous approaches have been developed to generate individual skyrmions or ensembles thereof, such as quasi-static magnetic field cycling \cite{moon2021universal}, magnetic field pulses \cite{MagneticFieldPulsesTheory,MagneticFieldPulsesExperiment}, and electronic nucleation \cite{BUTTNER2021255}. More recently, all-optical approaches have triggered interest \cite{PhysRevApplied.21.034065,zhu2024ultrafast,buttner2021observation}. They would provide a free-space method, that would allow for the free positioning of skyrmions, while also being significantly less restrictive on sample fabrication. Moreover, it opens up possibilities for magneto-photonic applications.

Several attempts to optically nucleate sub-micron skyrmions have been reported. Initially, research showed the possibility of stochastically generating an ensemble of skyrmions using single femtosecond laser pulses, exclusively with the addition of an external magnetic field \cite{buttner2021observation,zhang2023optical,ogawa2015ultrafast,je2018creation}. This was later followed by Zhu et al.\ \cite{zhu2024ultrafast}, who replicated the nucleation of random ensembles without the requirement of an external field. However, the possibility of not only optically writing a single sub-micron skyrmion deterministically, but being able to annihilate it deterministically as well, has remained an elusive goal. Note that the external magnetic field is both used for the generation of skyrmions and for the stabilization afterwards, resulting in a significantly restrictive environment.
\begin{figure}[H]
	\centering
	\includegraphics[width=0.99\linewidth]{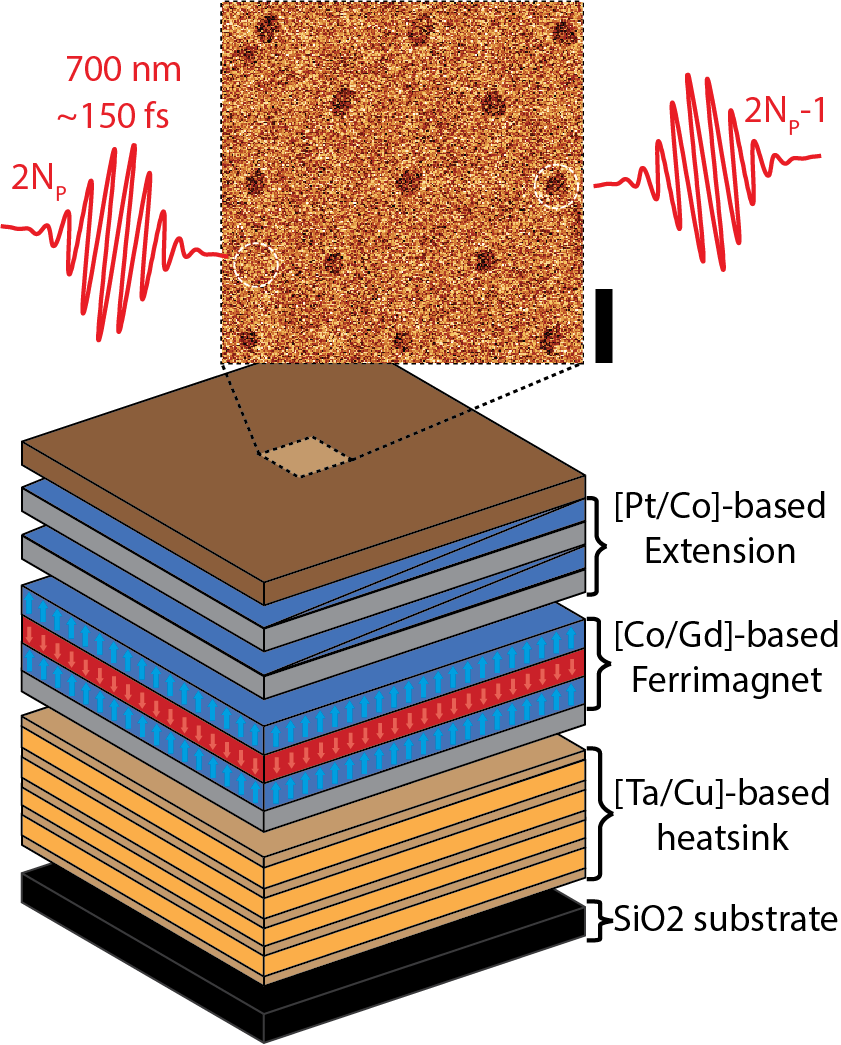}
	\caption*{Figure 1: Schematic illustration of the essence of the experiment performed in this research. At the bottom, a magnetic thin-film stack composed of 3 components: heatsink, a Co/Gd-based ferrimagnet for all-optical switching, and a Pt/Co-based extension. At the top, femtosecond laser pulses are directed at the sample, the inset showcasing an MFM image of a grid switched with said pulses. An uneven number of pulses ($2N_{\text{P}}-1$) results in a switched domain, whereas an even number of pulses ($2N_{\text{P}}$) brings it back to the original state}
\end{figure}
Here, we propose an alternative field-free method to stochastic nucleation, i.e., All-Optical Switching (AOS) of the magnetization using perpendicularly magnetized Co/Gd based ferrimagnetic stacks \cite{lalieu2017deterministic,PhysRevB.111.064421,davies2022helicity,nhjx-73gx}. In AOS, femtosecond laser pulses heat the electron gas temporarily above the Curie temperature, after which exchange-mediated transfer of spin angular momentum reverses the orientation of the magnetization in the two spin sub-lattices. Applied to skyrmion writing, the ultrafast laser heating leads to a quenching of the topological barrier, after which a transition can occur from the topologically trivial ferromagnetic state to a non-trivial chiral skyrmion state --- or vice versa --- within a few picoseconds. Due to the deterministic toggle process, it can be used to both generate and annihilate skyrmions. If energies are tuned slightly above the sharp switching (energy density) threshold, sub-diffraction limited domains can be written via this method. However, due to the ferrimagnetic nature of Co/Gd-based stacks, and thus weaker dipolar fields, the minimum stable domain diameter is usually on the scale of several microns. On top of this, not all Co/Gd based magnetic stacks guarantee significant DMI and thus chiral domains \cite{metternich2025defects}. Therefore, improvements are required to establish a method for the generation of sub-micron chiral skyrmions.

In this paper, we demonstrate deterministic writing and annihilation of individual sub-micron skyrmions using single femtosecond laser pulses in specially engineered Co/Gd-based magnetic thin films without the need for an external magnetic field. Our study resolves a competition between deterministic AOS and different stochastic processes as a function of sample composition, including layer thicknesses and the inclusion of additional Pt/Co repeats, capped with Ir, to both induce a larger magnetic moment and DMI while keeping switching properties intact. Also, we demonstrate the fixed chirality of the optically written Néel-type domains. Guided by this study, we find near optimal parameters for which we establish the sought for control of single sub-micron domains. Finally, we discuss the broader impact of our results, by which we open up a field-free avenue for exploring skyrmion-based systems and potential applications.

\section*{Results}
    \label{Sections:Results}
	\subsection*{Analyzing the switching processes}
The exact details and growth conditions of the magnetic stacks used in this research are given in the methodology. Each stack consists of 3 main components: A core Co/Gd/Co tri-layer, which provides the system with the property of AOS. On top of that, one or two repeats of Pt/Co are deposited and capped with Ir to provide a higher magnetic moment and larger DMI. The thickness of the Co layer in the additional Pt/Co repeat(s) $t_{\text{Co}}$ is varied using wedge growth. Finally, to control the heat dissipation, a Ta/Cu multi-layer is optionally included below the stack. A general schematic of such a magnetic stack is illustrated in Fig.\ 1. 
\begin{figure}[H]
	\centering
	\includegraphics[width=0.99\linewidth]{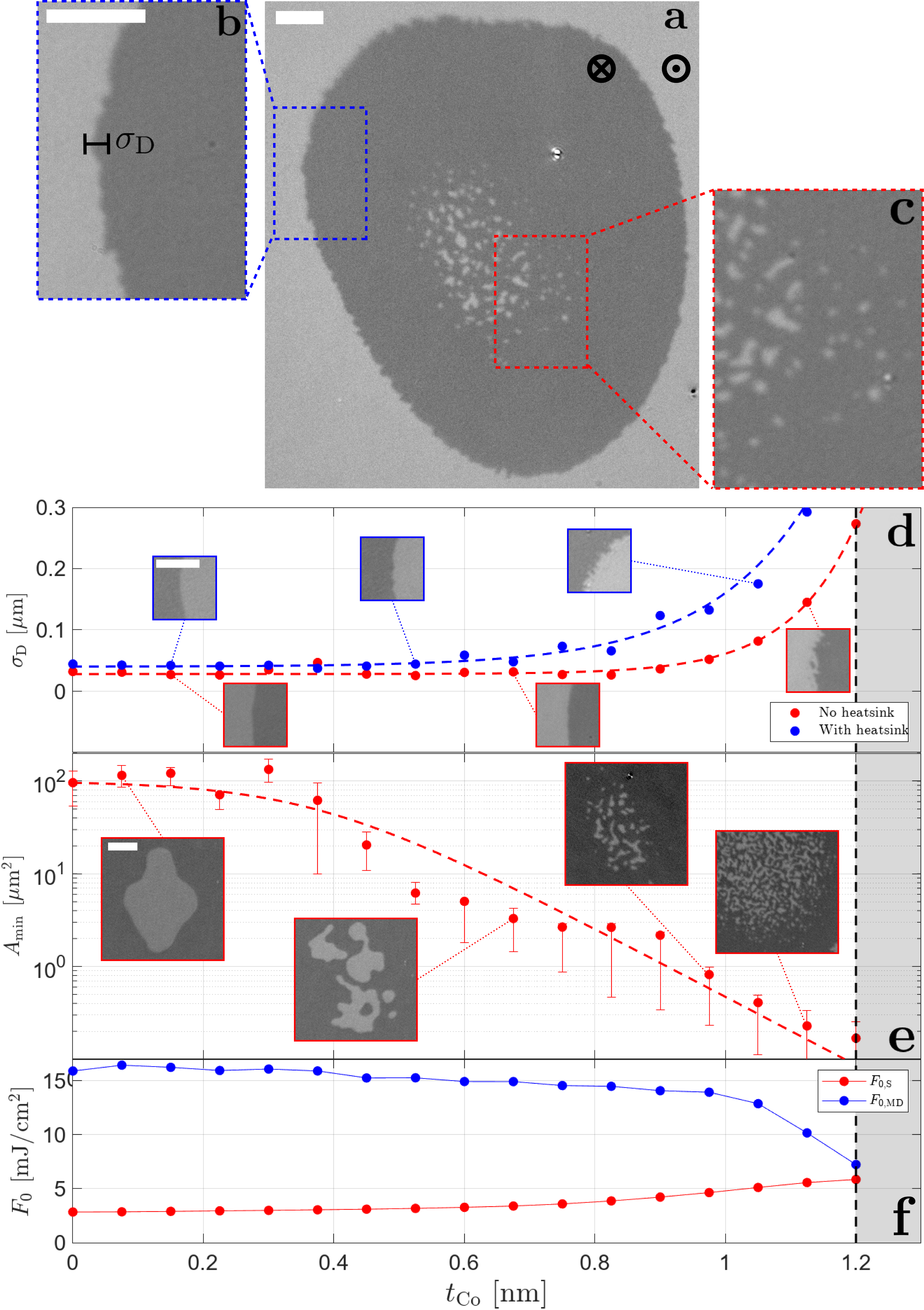}
	\caption*{Figure 2: Kerr microscopy and analysis of magnetic stack with a single Pt/Co extension with and without the inclusion of a heat sink. (a) Kerr image of a domain generated after a single femtosecond laser pulse for $t_{\text{Co}}=1.05$ nm and $E_p=264$ nJ. Light and dark contrast correspond to magnetization up (original direction) and down, respectively. (b-c) Two insets further illustrating the roughness $\sigma_D$ of the domain wall and the scale of the generated multidomain in (a). (d) $\sigma_D$ plotted against $t_{\text{Co}}$ for the magnetic stack with and without a heat sink, including a phenomenological fit ($a+be^{ct_{\text{Co}}}$) for guidance. (e) $A_{\text{min}}$ plotted against $t_{\text{Co}}$ without a heat sink, also including a phenomenological fit ($(a+be^{ct_{\text{Co}}})^{-1}$) for guidance. The errorbars represent the first and third quartile linked to each data point. Both (d) and (e) contain insets at different $t_{\text{Co}}$ for illustration. (f) Threshold fluence $F_0$ for AOS and multidomain formation for different $t_{\text{Co}}$. The grayed-out area for $t_{\text{Co}}>1.2$ nm indicates in-plane magnetization. All scalebars are 5 micron.}
\end{figure}
We first discuss the results for a magnetic stack with a single extension of Pt(1.25)/Co($t_{\text{Co}}$) without a heat sink, where $t_\text{Co}$ is varied between 0 and 1.5 nm. As a function of increasing cobalt layer thickness, the magnetic moment (and thus dipolar field strength) will increase, potentially leading to smaller stable domains. Moreover, the perpendicular anisotropy in the stack reduces, until it vanishes around 1.2 nm. This results in a transition from out-of-plane to in-plane magnetization, also known as the Spin Reorientation Transition (SRT). As beyond this thickness the magnetization turns in-plane, results are only shown for $t_{\text{Co}}\leq 1.2$ nm. In order to explore the optical switching process, individual 150 fs laser pulses of different pulse energy $E_{\text{p}}$ are focused to a spot size of 30-40 $\mu$m FWHM at different positions of the sample, corresponding to different $t_{\text{Co}}$. The resulting switched areas are imaged using a Kerr microscope. Due to the Gaussian energy distribution of each pulse, different local responses are recorded for a single pulse, depending on the local energy density or fluence $F$ [mJ/cm$^2$].

A Kerr microscopy image of the response to a single laser pulse for $t_{\text{Co}}=1.05$ nm is shown in Fig.\ 2a, where three different responses can be identified. Further away from the center of the laser spot, no switching occurs because the local fluence remained below the threshold, leaving the magnetic film in the original state (light contrast). Moving inwards, a homogeneously switched region is observed (dark contrast), as the local fluence has passed the threshold for AOS. At the center of the pulse, with the highest local fluence, a stochastic pattern of small back-switched domains (light contrast), with a local magnetization in the same direction as before the laser impact is observed. The multidomain region can be separated in two, where outer part consists of spontaneous stochastic generation of nearly circular sub-micron skyrmions, while the inner part contains a more labyrinth-like multidomain pattern. The skyrmion generation is only observed for $t_{\text{Co}}>0.8$ nm. We conjecture that their formation must have happened well after completion of the deterministic AOS process, and that it is a direct result of approaching the SRT \cite{zhang2023optical,liefferink2025effective}. A deeper analysis goes beyond the scope of this paper.

Figure 2f shows the threshold fluence for deterministic switching ($F_{\text{S}}$) and multidomain generation ($F_{\text{MD}}$) for different $t_{\text{Co}}$, as derived via the Liu method given in the methodology. The figure then illustrates a fluence gap ($F_{\text{MD}}-F_{\text{S}}$), indicating the fluence range in which one can deterministically switch the magnetization without multidomain generation. The fluence gap is found to reduce for higher $t_{\text{Co}}$, while it collapses at the SRT. Approaching this point, it becomes increasingly more important to have a more stable laser output with a small pulse-to-pulse energy fluctuation, which is typically a few \% for the laser system used in this work.

Zooming in on the outer contour of the AOS switched region in Fig.\ 2b, a clear meandering of the domain wall is observed. We attribute this to the reduced domain wall energy as the anisotropy is lowered when approaching the SRT. Shortly after the domain wall is formed --- during the cooling down after the laser heating --- a competition between gain in dipolar energy and cost of domain wall energy can result in such a thermally-assisted roughening. Being of stochastic character, it can undermine deterministic switching because of the mismatch between the switching contour of a next optical pulse and the exact shape of the previously written roughened domain. As a result, leftover domains will remain on the size scale of the meandering. We will later show that this same effect can also hinder the deterministic switching of sub-micron skyrmions. We quantify the meandering of the domain wall using a parameter $\sigma_D$, equal to the standard deviation from a least-squared fit line through the domain wall. The extracted values of $\sigma_{D}$ corresponding to our measurements for different $t_{\text{Co}}$ are shown in Fig.\ 2d. While it remains constant for a Co thickness up to 0.6 nm, for larger thicknesses it is found to grow exponentially when approaching the SRT. 

Shifting focus back to the multidomain generated in the center of the laser pulse, it should be noted that the domains stochastically generated here represent the minimum stable domain size, as that is the energetically most favorable state after complete demagnetization \cite{Coey_2010}. Therefore, extracting the domain size from these Kerr images allows us to estimate the minimum domain size $A_{\text{min}}$ for different $t_{\text{Co}}$ as illustrated in Fig.\ 2e. While initially $A_{\text{min}}$ remains constant, for larger Co thicknesses, $A_{\text{min}}$ decreases exponentially.

The observed trend implies that maximizing $t_{\text{Co}}$ should lead to the smallest possible stable domains. However, if $t_{\text{Co}}$ is increased too much, stochastic processes like the meandering of the domain walls, multidomain generation, or spontaneous skyrmion generation will dominate the dynamics. This will eventually rid the system of deterministic toggle switching, especially at the sub-micron scale.

An important property of skyrmions is their fixed chirality set by the net DMI, and their topological charge of $\pm$1. To verify the chirality of domain walls in our system, we performed Scanning Electron Microscopy with Polarization Analysis (SEMPA) to determine the out-of-plane and in-plane orientation of the magnetization in the domains and at the center of the walls, respectively, as further described in the methodology. Since SEMPA is sensitive to the top nanometers only, we did so for a magnetic stack identical to the one used in this chapter for $t_{\text{Co}}=$ 0.6 nm, but without an Ir(2)/Ta(4) capping. Fig.\ 3a illustrates such a SEMPA measurements of a labyrinth-like multidomain generated via a femtosecond laser pulse.
\begin{figure}[H]
	\centering
	\includegraphics[width=0.99\linewidth]{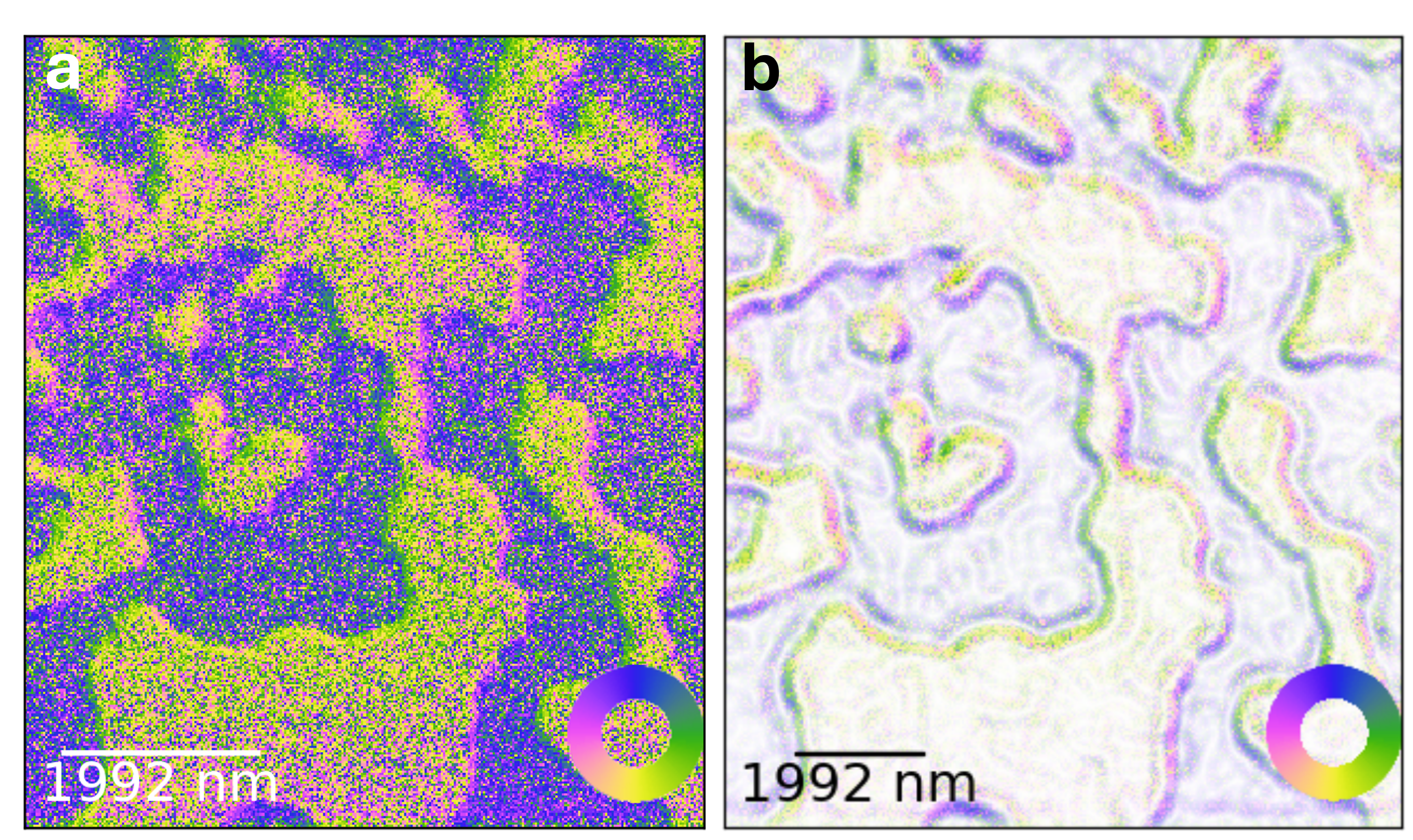}
	\caption*{Figure 3: SEMPA images of a magnetic stack with a single Pt/Co extension without a heat sink. (a) raw SEMPA image where up-magnetization is indicated in yellow, while down-magnetization is indicated in blue, superposed on the in-plane contrast as indicated by the color wheel. (b) SEMPA image of the same domain with further data processing for clearer visualization of the domain wall.}
\end{figure}
In this image, the yellow domains indicate up-magnetization, whereas the darker blue domains indicate down-magnetization. Through further image processing we can extract the in-plane magnetization in the domain wall as illustrated in Fig.\ 3b \cite{ThesisMarkDeJong}. Here, the colour of the domain wall directly correlates with the local in-plane spin orientation via the colour wheel at the bottom right. From this, we can directly conclude that our magnetic stacks have Néel-like domain walls with well-defined counter-clockwise chirality. We emphasize that even without a top Ir cap, the effective DMI is strong enough to stabilize domains into a Néel texture with fixed chirality. Furthermore, the chirality of a given domain was found not to depend on the nature of how it was generated (i.e., the same handedness for the meandering of domains, sub-micron stochastic skyrmions and the outer contour of the AOS switched region was observed; See S.I.), but is instead an intrinsic property of the magnetic stack itself.

\subsection*{Writing and annihilating single skyrmions}\label{Sec:4.3}
Equipped with the insights on stochastic and deterministic dynamics following exposure by single fs laser pulses, we continue with exploring the deterministic switching of sub-micron skyrmions. It is emphasized that all these results are obtained without a stabilizing applied magnetic field. We use a high NA lens described in the methodology, producing a laser spot size of 1 $\mu$m FWHM. As is also described in the methodology, the domain size of a switched domain not only scales with the laser spot size, but also with the pulse energy relative to the fluence threshold. This allows us to switch domains that are well below the diffraction limit.

For these experiments we use samples with a constant Co thickness, the optional inclusion of a heat sink, and either a single or a double Pt(1.25)/Co($t_{\text{Co}}$) extension. For each sample, and for a larger number of different pulse energies, we generate a grid of pulse sequences, where at each position a different number of pulses is fired, ranging from 1-10, with a temporal gap of 10 ms. If deterministic switching is present, an uneven number of pulses will result in a net reversal of the original state, while an even number of pulses does not. An example of a full dataset can be found in S.I.2.
\begin{figure*}[!tp]
	\centering
	\includegraphics[width=\linewidth]{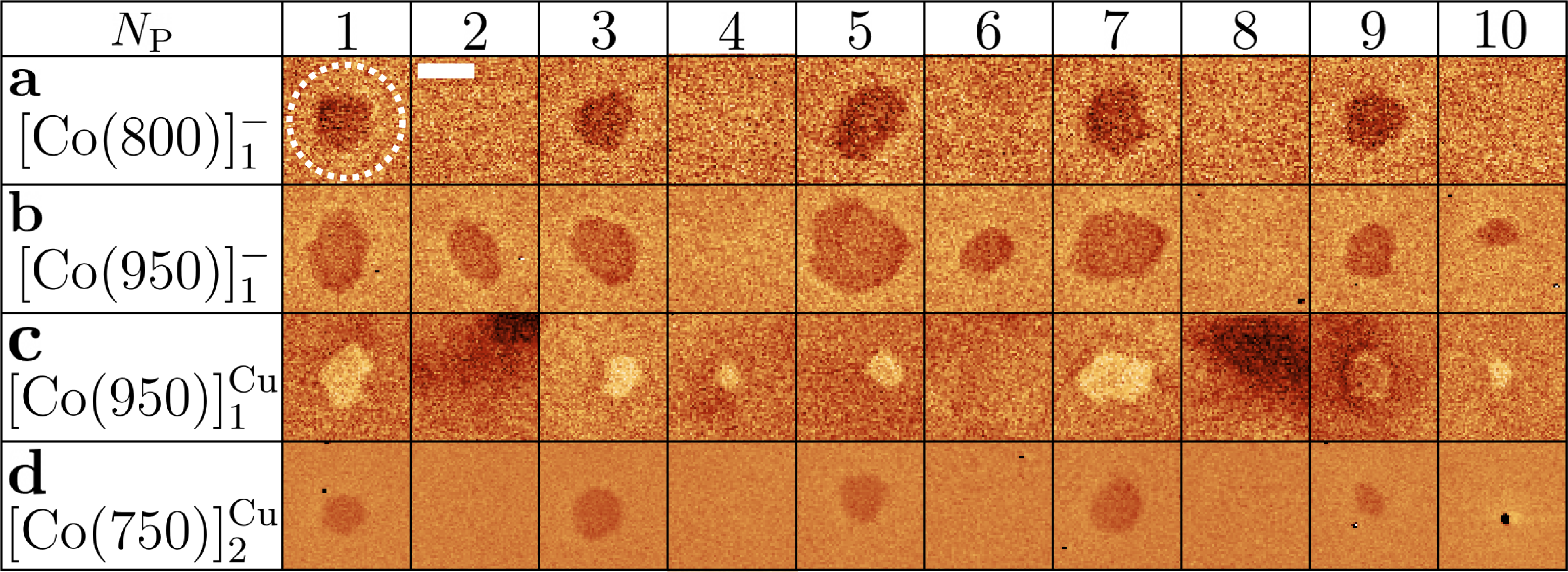}
	\caption*{Figure 4: MFM images illustrating the result of firing $N_\text{P}$ laser pulses at different stack compositions. For the compositions, the thickness of the Co layer is indicated in picometers in brackets and the number next to the square brackets indicates the number of Pt(1.25)/Co($t_{\text{Co}}$) extensions. The addition of "Cu" next to the square brackets indicates the inclusion of a heat sink. The pulse energies for each composition are 43, 69, 78 and 89 pJ, respectively. The dotted white circle represents the laser spot size of 1 $\mu$m. The scalebar is 500 nm.}
	\label{Fig:SkyrmionSwitching}
\end{figure*}
A summary of the switching result for four different sample compositions --- each attained slightly above the threshold fluence --- is given in Fig.\ 4. The sample with a single Pt/Co repeat with $t_{\text{Co}}=0.8$ nm and without a heat sink is shown in Fig.\ 4a. Clear deterministic toggle switching is observed, the smallest domains generated having diameters of approximately 400 nm. However, it was also observed that the smallest domains would spontaneously annihilate over the span of days, not providing long-term stability. Increasing $t_{\text{Co}}$ to 0.95 nm (Fig.\ 4b) further increases the stability of generated domains, bringing the stable diameter down to 200 nm but decreasing the power range in which one can switch (See Fig.\ 2f). The result is that even slightly above the threshold fluence, consistent deterministic toggle switching is no longer achieved. Although, after an even number of pulses a domain consistently shows up, after an odd number of pulses annihilation is not complete, leaving a remaining smaller domain. We attribute this to laser pulse-to-pulse fluctuations in combination with stochastic thermal processes from the reduced fluence range, as well as large stochastic effects (meandering) for larger $t_{\text{Co}}$.

To combat the thermal processes, a heat sink was added to the bottom of the magnetic stack, analogous to AOS studies by Verges et al.\cite{PhysRevApplied.21.044003} They showed that including a heat sink significantly increases the threshold fluence of multidomain generation without affecting the AOS threshold fluence. We confirm this behaviour, but also observe a significant increase in the meandering $\sigma_\text{D}$ as shown in the top graph of Fig.\ 2d.

Results for sub-micron writing for the $t_{\text{Co}}=0.95$ nm stack with heat sink are displayed in Fig.\ 4c. In line with the stronger meandering, the sub-micron domains show a less regular shape, which ends up being detrimental to deterministic switching. As a consequence, leftover domains are still observed for an even number of switches for this stack. A solution is provided in the form of a double Pt(1.25)/Co(0.75) repeat, for which results are shown in Fig.\ 4d. The extra Pt layer increases the perpendicular anisotropy (due to the additional Pt/Co interfaces), but also reduces the minimum stable domain size (due to the increased magnetic moment). This finally restores the property of deterministic magnetic switching, generating domains with diameters down to 175 nm, observed to be stable over months. On top of that, due to the addition of the heat sink, $F_{\text{MD}}$ was observed to be approximately three times larger than $F_{\text{S}}$, providing a large range of pulse energies in which one can toggle switch magnetic domains. Firstly, this reduces the effect of pulse-to-pulse energy fluctuations. Secondly, it also allows for the switching of skyrmions of significantly different sizes ranging from the minimum of 175 nm to approximately 1 $\mu$m before reaching stochastic processes (See S.I.1).

Finally, we want to address the stability of our nanometer-scale skyrmions without applying a stabilizing magnetic field, whereas most other work relies on such a field. It is known that the relation between skyrmion energy and its radius describes the existence of two energy minima, due to different mechanisms and in different skyrmion diameter regimes \cite{buttner2018theory}. The first minimum corresponds with a 'true' skyrmion stabilized by DMI, with a diameter of typically 10 nm and a reversed magnetization only at a singular point in its center. The second energy minimum is only found in an applied magnetic field that opposes the spins at the center of the skyrmion, and corresponds to a bubble skyrmion --- larger reversed domains surrounded by a chiral domain wall --- with similar diameters as the ones observed in our work. We propose that our skyrmions are meta-stable due to their presence near the energy maximum in the total energy function radius of the skyrmionic texture, located between the regimes of DMI- and field-stabilized skyrmions. Driving fields to let domains collapse or expand are linked to the derivative of the skyrmion energy. As a consequence, near the energy maximum, these fields are smaller than typical domain wall depinning fields, making skyrmions quasi-stable for at least months.

\section*{Discussion}
    \label{Sections:Discussion}
	In our work we highlighted the competition between deterministic skyrmion formation due to AOS, and stochastic processes. Therefore, understanding the latter is relevant to improve purely deterministic control. Optically-induced generation of random ensembles of skyrmions in small magnetic fields has been reported by Büttner et al., as explained by a fluctuation-driven mechanism upon ultrafast laser heating above the Curie temperature \cite{buttner2021observation}. Our results clearly show formation of stochastic skyrmion ensembles --- next to labyrinth-like multidomain generation --- even in zero magnetic field, when approaching the SRT for thicker Co thickness. For this peculiar find, we refer to Zhang et al.\ \cite{zhang2023optical}. They state that, as a magnetic stack in the SRT is heated up, it will temporarily transition from an out-of-plane to an in-plane magnetization. As the system cools down it regains its perpendicular anisotropy. With the addition of an external field, they show that this can result in stochastic skyrmion generation. A key difference in our findings is that we do not require an external field to replicate this result. It should also be noted that it is not necessarily a pure heating effect, like labyrinth-like multidomain generation. The inclusion of a heat sink in our magnetic stack did increase the fluence threshold for labyrinth-like generation significantly, but did not do so for the stochastic skyrmion ensemble nucleation. In the end, for our deterministic purposes, one can avoid these stochastic processes by maintaining a high enough perpendicular anisotropy.

There is also ample room for stability and domain size improvement. Note that no rigorous testing was done for how extensive the Pt/Co extension can be. Nor was testing done for different Pt thicknesses in the extension layer. Due to the exponential nature of decrease in stable domain size, further optimizations could potentially result in stable skyrmion diameters below 100 nm.

Our result opens up numerous possibilities within research on topological textures. Due to the deterministic nature and the flexibility of free-space AOS, any arbitrary lattice of skyrmions can be written at ones will. Moreover, our method of deterministic toggle switching of skyrmions also provides routes for the generation of more topologically complex structures. For example, by firing 2 consecutive pulses with different energies at the same position, a skyrmion bag of the 1st order --- or skyrmionium --- can be generated. 

In conclusion, using the process of all-optical switching by single fs laser pulses applied to specially tuned Co/Gd layer materials, we successfully demonstrated deterministic creation and annihilation of magnetic skyrmions with diameters down to 175 nm. Attaining this goal relied on a rigorous analysis of the competition between deterministic switching and different stochastic processes. Increasing the magnetic moment per area by depositing additional ferromagnetic material leads to a growing importance of stochastic labyrinth textures, spontaneous nucleation of skyrmions and meandering of domain walls of optically switched domains. The dominance of these effects appears to be strongly linked to the reduced magnetic anisotropy as well as the associated proximity of the spin reorientation transition. A thorough understanding of these effects is a crucial component to enable reliable optical control of nanometer-scale skyrmionic domains. We envision numerous applications of our findings in skyrmion-based research, and foresee unique possibilities to further scale down the size of the textures and exploit the same optical method to deterministically generate structures with more complex topology.
	
\section*{Methodology}
    \label{Sections:Methodology}
    \subsubsection*{Magnetic stack composition and growth}
All magnetic stacks used in this paper were grown via magnetron sputtering at a base pressure smaller than $1\cdot10^{-8}$ mbar on a substrate of Si of 500 $\mu$m capped with a 100 nm thick layer of SiO$_2$ formed via thermal evaporation. The varying thickness across a magnetic stack was achieved via the implementation of a moving wedge during the sputtering process. 

All magnetic stacks are built up from three components as illustrated in Fig.\ 1. Each stack contains a core built up of Ta(4)/Pt(4)/Co(0.6)/Gd(1.2)/Co(0.6) for AOS, where all thicknesses are given in nanometers. Some stacks also include a heat sink below said core, which consists of [Ta(2)/Cu(10)]$_{10}$. On top of the core is a Pt/Co based extension, which either consists of a single layer or a double layer of Pt(1.25)/Co($t_{\text{Co}}$). Finally, all layers are capped with Ir(2)/Ta(4). An example of a full double extension stack with a heatsink thus goes as follows: [Ta(2)/Cu(10)]$_{10}$/Ta(4)/Pt(4)/Co(0.6)/Gd(1.2)/- Co(0.6)/[Pt(1.25)/Co($t_\text{Co}$)]$_2$/Ir(2)/Ta(4). For SEMPA measurements we deviate from this design. Because of its sensitivity to only the top few nanometers, the Ir/Ta cap is left out. Those samples are stored and measured at ultra-high vacuum (10$^{-10}$ mbar or below).

\subsubsection*{Laser setup}
A Spectra-Physics Spirit-NOPA femtosecond laser has been used, providing an output of 700 nm laser pulses with a temporal length of 150 fs. Experiments with larger spot sizes were achieved via the use of standard optics, while the sub-micron spot sizes were achieved using a Mitutoyo M Plan APO NIR HR 100x (0.70 NA) objective. The pulse energy $E_{\text{p}}$ was varied using a neutral density filter wheel.

\subsubsection*{Kerr microscopy}
Kerr microscopy images were made using a Kerr microscope from Evico magnetics. Basic post-processing methods such as background subtraction were applied to enhance the magnetic contrast.

\subsubsection*{Magnetic force microscopy}
Magnetic Force Microscopy (MFM) images were made using a Bruker Dimension Edge AFM setup. MFM tips were home-made by sputtering the following magnetic stack onto AFM tips: Ta(4)/Co(5)/Ta(4). The magnetron sputtering settings are identical to the ones used to sputter the magnetic stacks. 

\subsubsection*{SEMPA}
SEMPA images were made using a ScientaOmicron SEMPA lab construct containing a SEM from Zeiss and an added Spin Polarized Low Electron Energy Diffraction (SPLEED) from FOCUS for the added polarization analysis. In standard operation, SEMPA provides (vectorial) in-plane magnetic contrast only, therefore, to measure the out-of-plane magnetization of our magnetic stacks, the setup is tilted by 5 degrees relative to the SPLEED detector. This superposes the out-of-plane magnetization onto the in-plane y-direction, resulting in a yellow color for up-magnetizaton and a blue color for down-magnetization after image processing.

\subsubsection*{Calculating the fluence threshold}
Using the Liu method \cite{liu1982simple} one can determine the threshold fluence $F_0$ with the following equation:
\begin{equation*}
	A = \pi r \sigma ^2 \ln (\frac{E_p}{\pi r \sigma^2F_0}),
\end{equation*}
where $A$ is the area of the affected domain, $E_p$ the pulse energy, $F_0$ the threshold fluence, $\sigma$ the beam radius along the short axis of a slightly elliptical Gaussian pulse, and $r$ the ratio between the beam radius along the long and short axis.

\section*{Acknowledgments}
	\label{Sections:Acknowledgments}
	We thank Julian Hintermayr for designing and constructing the optical setup that enabled SEMPA measurements. We acknowledge the research program “Materials for the Quantum Age” (QuMat) for financial support. This program (registration number 024.005.006) is part of the Gravitation program financed by the Dutch Ministry of Education, Culture and Science (OCW).
\end{multicols}

\clearpage
\printbibliography
\clearpage

\section*{Supplementary Information}
	\label{Sections:Supplementary}
	\subsection*{Full dataset for writing and annihilating single skyrmions}
To switch and visualize single skyrmions, magnetic stacks were grown in 50$\times$50 $\mu$m islands constructed via electron beam lithography. One full dataset generated on three different islands for the magnetic stack with a double Pt(1.25)/Co(0.75) repeat is shown in Fig.\ S.I.1. For each fluence a 5$\times$2 grid was switched, where each position in the grid contained an additional laser pulse, with an internal spacing of 2 micron between each position (See S.I.1d). This structure was then repeated in a 3$\times$10 grid over an entire island, where each structure contained a different laser pulse energy. 
\begin{figure}[H]
	\centering
	\includegraphics[width=\linewidth]{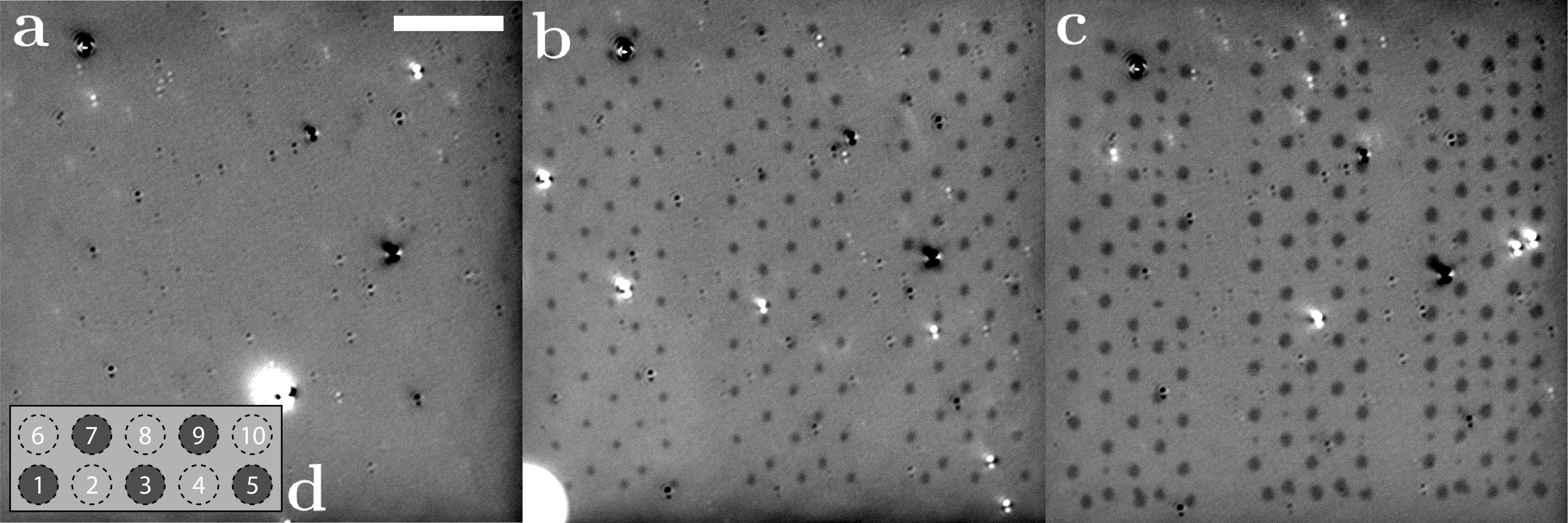}
	\caption*{Figure S.I.1: Kerr images of the full dataset of switching single skyrmions on the magnetic stack with a double Pt(1.25)/Co(0.75) repeat. (a), (b) and (c) contain laser pulse energies varying from 3.2-94.9, 98.1-189.8 and 193.0-284.7 pJ, respectively. The lighter grey indicates the initial magnetization state, whereas the darker grey represents the magnetization opposite to the initial state. (d) contains a schematical illustration of a single 2$\times$5 grid of pulses, which a number indicating the number of pulses at each position. The spacing between each point in both the x- and y-direction is 2 $\mu$m. The scale bar is 8 $\mu$m.}
	\label{Fig:SupplSwitchData}
\end{figure}
As not all data fit into a single 50$\times$50 $\mu$m island, multiple islands were used. The laser pulse energy was set to 3.2 pJ for the first set of switches, and then incrementally increased by 3.2 pJ per set. The full energy range used was from 3.2 to 284.7 pJ, where the first, second and third island used contained energies ranging from 3.2-94.9, 98.1-189.8 and 193.0-284.7 pJ, respectively. Each Kerr image then shows the full dataset for the first, second and third island in Fig.\ S.I.1a, S.I.1b and S.I.1c, respectively. The set of switches used in Fig.\ 4d can be vaguely seen in the top right of Fig.\ S.I.1a.
 
\subsection*{Additional SEMPA images}
The chirality of the domain walls was tested for different methods of domain formation. In addition to Fig.\ 3, which illustrates the chirality for a labyrinth-like multidomain generated with a femtosecond laser pulse, Fig.\ S.I.2 does so for a domain wall generated via AOS (FIG.\ S.I.2(a-b)) and a multidomain that formed as grown (Fig.\ S.I.2(c-d)) on a magnetic stack with a single extension of Pt(1.25)/Co(0.6) without Ir(2)/Ta(4) capping.
\begin{figure}[H]
	\centering
	\includegraphics[width=\linewidth]{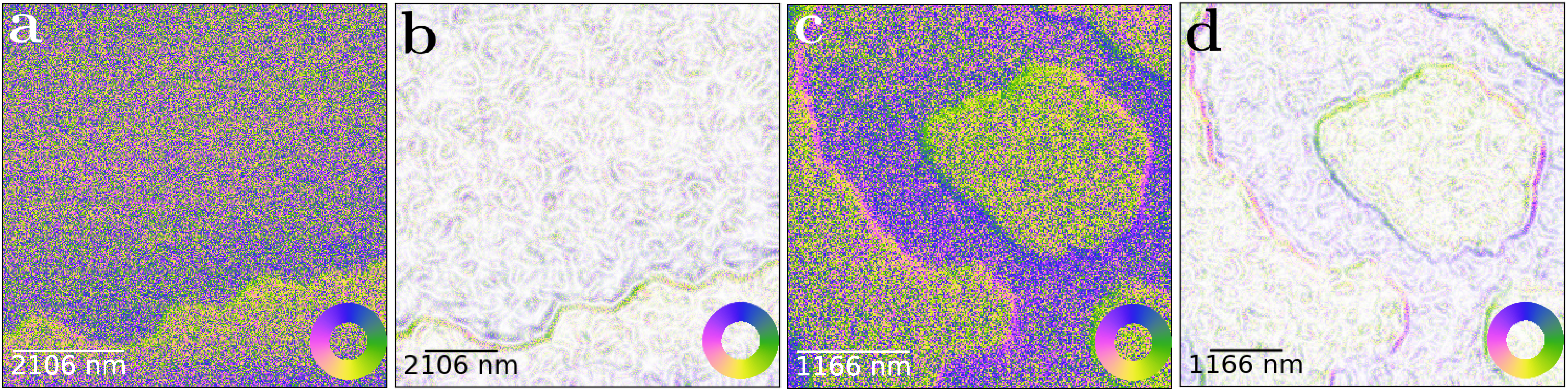}
	\caption*{Figure S.I.2: SEMA images of differently generated chiral domain walls on a magnetic stack with a single extension of Pt(1.25)/Co(0.6). (a) and (b) show the raw SEMPA data and further processed data with a clearer domain wall for a transition from up- (yellow) to down-magnetization (blue) generated with a femtosecond laser pulse, respectively. (c) and (d) illustrate the same data for a multidomain generated on the magnetic stack as grown. The spin direction inside the domain walls is illustrated by the color wheel.}
	\label{Fig:SupplSEMPAData}
\end{figure}
All observed domain walls are of the same anti-clockwise Néel-like nature as the domain walls observed in Fig.\ 3. From this it has been concluded that the nature of the generation of the magnetic domains and thus domain walls is not critical to the domain wall type generated, but is instead an intrinsic property of the stack.
	
\end{document}